\documentclass[journal=jctc,manuscript=article]{achemso}

\usepackage[T1]{fontenc} 
\usepackage{tgpagella}   
\usepackage{microtype}   
\usepackage{hyperref}    
\usepackage{fontawesome} 

\usepackage{amsmath}    
\usepackage{mathtools}  
\usepackage{bm}         
\usepackage{physics}    
\usepackage{tensor}     
\usepackage{braket}     
\usepackage{booktabs}   

\usepackage[version=3]{mhchem} 

\usepackage{graphicx} 
\usepackage{dcolumn}  
\usepackage{multicol} 
\usepackage{subcaption}
\usepackage{float}

\usepackage{verbatim} 
\usepackage[x11names]{xcolor} 
\usepackage{soul}     

\usepackage{bbold}    
\usepackage{systeme}  

\usepackage{pdfpages}

\allowdisplaybreaks[1]
\graphicspath{{}}

\SectionNumbersOn

\author{Alberto Barlini}
\affiliation{Scuola Normale Superiore, Pisa, Italy}

\author{Andrea Bianchi}
\affiliation{Scuola Normale Superiore, Pisa, Italy}

\author{Enrico Ronca}
\email{enrico.ronca@unipg.it}
\affiliation{Dipartimento di Chimica, Biologia e Biotecnologie, Università degli Studi di Perugia, Perugia, Italy}

\author{Henrik Koch}
\email{henrik.koch@ntnu.no}
\affiliation{Department of Chemistry, Norwegian University of Science and Technology, Trondheim, Norway}

\title[]{Theory of magnetic properties in QED environments: application to molecular aromaticity}

\abbreviations{ab initio QED, polaritonic chemistry, QED response theory, strong coupling, magnetic properties}


\begin{document}


\begin{abstract}
    In this work, we present \textit{ab initio} cavity QED methods which include interactions with a static magnetic field and nuclear spin degrees of freedom using different treatments of the quantum electromagnetic field. We derive explicit expressions for QED-HF magnetizability, nuclear shielding, and spin-spin coupling tensors. We apply this theory to explore the influence of the cavity field on the magnetizability of saturated, unsaturated, and aromatic hydrocarbons, showing the effects of different polarization orientations and coupling strengths. We also examine how the cavity affects aromaticity descriptors, such as the nucleus-independent chemical shift and magnetizability exaltation. We employ these descriptors to study the trimerization reaction of acetylene to benzene. We show how the optical cavity induces modifications in the aromatic character of the transition state leading to variations in the activation energy of the reaction. Our findings shed light on the effects induced by the cavity on magnetic properties, especially in the context of aromatic molecules, providing valuable insights into understanding the interplay between the quantum electromagnetic field and molecules.
\end{abstract}

\begin{figure}[ht]
    \centering
    \includegraphics[scale=0.6]{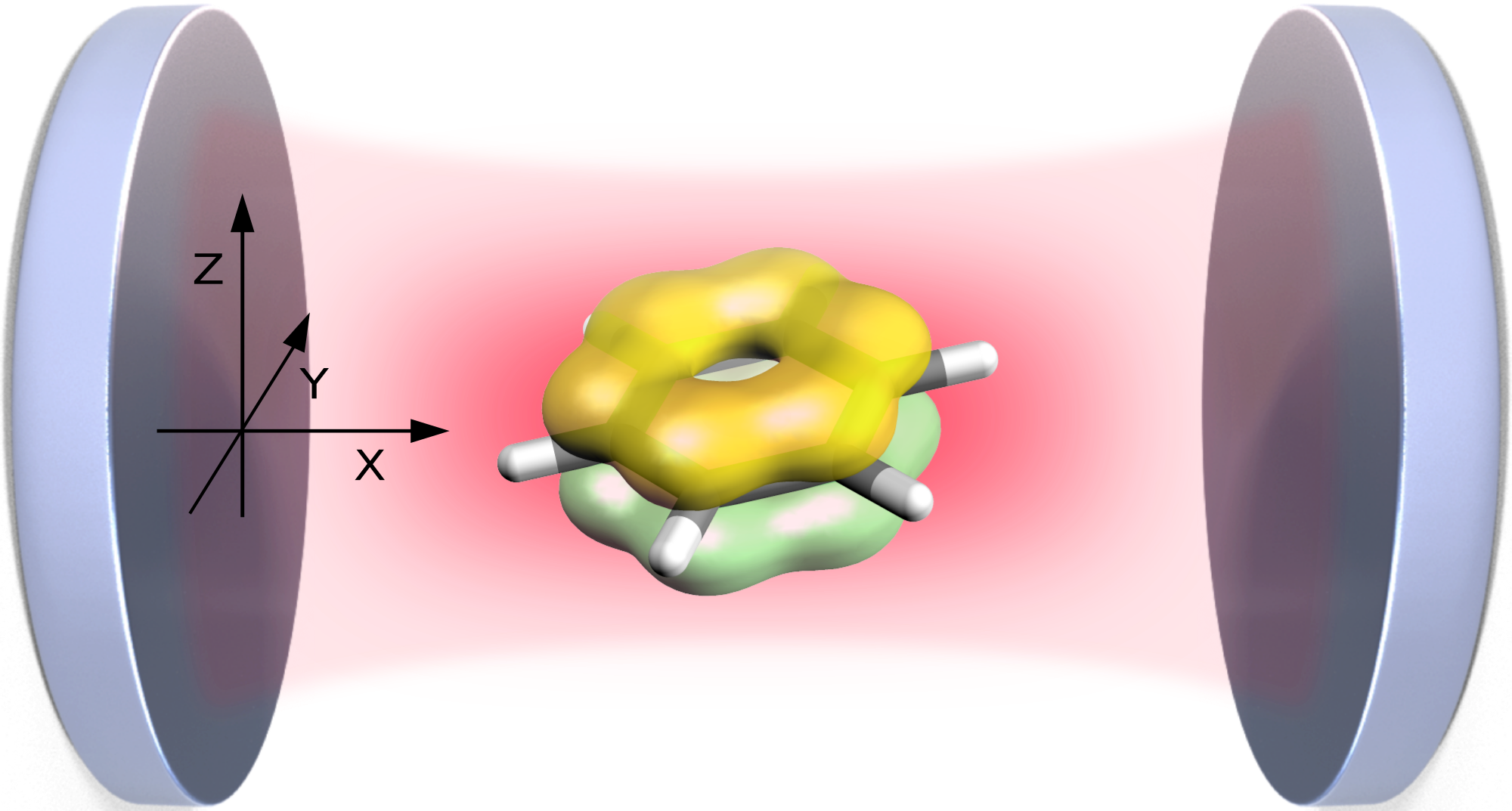}
    \caption{Graphical representation of the benzene molecule in an optical cavity.}
    \label{fig:opt_cav}
\end{figure}

\section{Introduction}
Polaritonic chemistry has recently gained significant attention, thanks to pioneering research by Ebbesen et al. \cite{hutchison2012modifying}, which demonstrated that the strong-light matter coupling can influence photochemical reactions and ground state reactivity \cite{thomas2016ground,lather2019cavity,thomas2019tilting}. Several experimental findings have revealed the influence of electromagnetic confinement on a wide range of processes, including chemical reactions \cite{hutchison2012modifying,thomas2016ground,lather2019cavity,thomas2019tilting,munkhbat2018suppression,canaguier2013thermodynamics,sau2021modifying,hirai2020modulation,ahn2023modification,hirai2022autotuning}, singlet fission \cite{eizner2019inverting, takahashi2019singlet, martinez2018polariton}, intersystem crossing \cite{stranius2018selective,yu2021barrier,ulusoy2019modifying}, and crystallization \cite{joseph2021supramolecular,hirai2021selective,sandeep2022manipulating}, as well as optical properties such as absorption, scattering, and emission \cite{garcia2021manipulating,chervy2018vibro,george2015ultra,xue2018ultrastrong,del2015signatures,baranov2020circular,guo2021optical,itoh2018reproduction,herrera2017absorption,wang2020coherent,takele2021scouting,barachati2018tunable,mund2020optical,ebadian2017extending,wang2014quantum,wang2021large,wright2023rovibrational}. \par  
Recently, experimental works have reported the effect of a quantum electromagnetic field on molecular magnetic properties. Eddins et al. \cite{eddins2014} reported the strong coupling of molecular nanomagnets within a microwave cavity. Ghirri et al. \cite{ghirri2015} developed devices that operate in the microwave range in the presence of strong magnetic fields. These devices have been used to couple photon and electronic spin degrees of freedom, showing potential applications in quantum information \cite{bonizzoni2017}. Jenkins et al.  \cite{jenkins2016} proposed a magnetic quantum processor composed of individual molecular spins coupled to superconducting coplanar resonators. Not only the field effects on the magnetic properties of matter have been investigated. Recently Ebbesen et al. \cite{ebbesen_nmr2023} demonstrated that standard nuclear magnetic resonance (NMR) spectroscopy is a viable tool to investigate vibrational strong coupling (VSC) effects inside microfluidic optical cavities. \par 
From a theoretical perspective, extensive progress has been made in recent years to describe the physical states of strongly light-matter coupled systems. Various \textit{ab initio} quantum electrodynamics (QED) approaches have emerged, including QED density functional theory (QEDFT) \cite{ruggenthaler2014quantum,tokatly2013time}, QED Hartree–Fock (QED-HF) \cite{Tor2020,haugland2021intermolecular}, strong coupling QED Hartree–Fock (SC-QED-HF) \cite{riso2022molecular}, second-order QED M\o ller Plesset perturbation theory (QED-MP2) \cite{bauer2023perturbation}, QED coupled cluster (QED-CC) \cite{Tor2020}, QED full configuration interaction (QED-FCI) \cite{Tor2020}, and more \cite{flick_2017,angelico_2023}. Recently, Rokaj et al. \cite{rubio_qed_mag_2019} proposed a theory for describing the interaction of solid-state materials coupled to a quantum electromagnetic field and a static external magnetic field of arbitrary strength. However, there are currently no theoretical studies on the quantum field effects on the magnetic properties of molecules. These properties involve magnetizability, defined as the second derivative of the energy with respect to an external magnetic field \cite{ruud1993hartree}, nuclear shielding, and indirect spin-spin couplings, both of which play a key role in simulations of NMR spectroscopy \cite{helgaker1999}. Moreover, both magnetizability and nuclear shielding tensors, are employed as aromaticity descriptors for molecules \cite{lazzeretti2004, deproft2002, poranne2015} and even aromatic transition states \cite{remco_trime_2003, jiao_aroma_1998, jiao_electrostatic_au_1995, jiao_nmr_trime_1993, jiao_electro_1994, jiao_magsus_trime_1994, jiao_mag_trim_1995}. Specifically, the nucleus-independent chemical shift (NICS) serves as a quantitative and qualitative gauge of the induced magnetic field within a molecule in an external magnetic field \cite{stanger_nics_2020}. In addition, the magnetizability exaltation quantifies the increase in magnetizability due to the electron delocalization associated with ring currents \cite{dauben_1968}. \par 
In this paper, we developed \textit{ab initio} methods to investigate quantum field-induced magnetic properties. In the first part of the paper, a general theory based on the minimal coupling Hamiltonian is presented. In Section \ref{cav_app}, the Hamiltonian with an approximate description of the cavity field that extends beyond the dipole approximation is derived. Starting from this general formulation we introduced the dipole approximation and derived the length gauge Hamiltonian, as presented in Section \ref{dip_H}. This Hamiltonian is used to derive a QED-HF approach to simulate magnetizabilities, nuclear shieldings, and indirect spin-spin couplings. The response formalism used to calculate these properties has been detailed in Section \ref{qed_hf_mag_prop}. In the last part of the paper, we apply our implementation to investigate the effects of the quantum field. In Section \ref{mag_hydro}, we present the results for the magnetizabilities of saturated, unsaturated, and aromatic hydrocarbons. In Section \ref{trim_cav}, we report the calculations of NICS and magnetizability exaltation. Finally in Section \ref{conclusions}, we present our concluding remarks.

\section{Theory}
In the upcoming sections, we will start from a QED minimal coupling Hamiltonian in the presence of a static magnetic field \cite{rubio_qed_mag_2019}. Then, we will derive a QED Hamiltonian with an approximated cavity field that goes beyond the dipole approximation. We will formulate the dipolar Hamiltonian and we will report its derivatives. Lastly, we will derive expressions for the QED-HF magnetizability, nuclear shieldings, indirect spin-spin couplings, and their response equations.

\subsection{QED Hamiltonian with a static magnetic field}
In the Born-Oppenheimer approximation, the radiation-matter interaction can be described in the non-relativistic limit by the minimal coupling Hamiltonian \cite{craig1998molecular}, which in atomic units reads as
\begin{gather} \label{H_minimal_coupling}
    \begin{split}
        {H}_{mc} & =\frac{1}{2} \sum_{i } \bm{\pi}_i^{2} + V - \sum_{i} \mathbf{m}_i \cdot \mathbf{B}\left( \mathbf{r}_i \right) + \frac{1}{8 \pi} \int \left( {\textbf{E}(\textbf{r})} ^2 + c^2  {\textbf{B}(\textbf{r})}^2 \right) d^3 \textbf{r} 
    \end{split},
\end{gather}
where the kinetic momentum operator $\bm{\pi}_i$ at the position of the electron \textit{i} is
\begin{gather} \label{kinetic_momentum}
    \bm{\pi}_i = \mathbf{p}_i + \mathbf{A} \left( \mathbf{r}_i \right).
\end{gather}
In Eq. \ref{kinetic_momentum}, $\mathbf{p}_i$ is the momentum operator, and $\mathbf{A} \left( \mathbf{r}_i \right)$ is the vector potential associated with the quantum electromagnetic field
\begin{gather} 
\begin{split}
    \mathbf{A} \left( \mathbf{r}_{i} \right) & = \sum_{\mathbf{k} \lambda} \mathcal{A}_{\mathbf{k}} \bm{\epsilon}_{\lambda} \left(  b_{\mathbf{k} \lambda} e^{i \mathbf{k} \cdot \mathbf{r}_{i}} + b_{\mathbf{k} \lambda}^{\dagger} e^{- i \mathbf{k} \cdot \mathbf{r}_{i}} \right)
\end{split}.
\end{gather}
The operators $b_{\mathbf{k} \lambda}^{\dagger}$ and $b_{\mathbf{k} \lambda}$ create and annihilate a photon with frequency $\omega_{\mathbf{k}}$, wave vector $\mathbf{k}$, and polarization $\bm{\epsilon}_{\lambda}$, respectively. The coupling strength is
\begin{gather}
    \mathcal{A}_{\mathbf{k}} = \sqrt{\frac{ 2 \pi}{ \varepsilon_{r} \omega_{\mathbf{k}} V_{\mathbf{k}}}}
\end{gather}
where the vacuum permittivity is equal to $1/4 \pi$ in atomic units, $\varepsilon_{r}$ is the relative permittivity, and $V_{\mathbf{k}}$ denotes the quantization volume of the mode defined by the wave vector $\mathbf{k}$. In Eq. \ref{H_minimal_coupling}, the electron magnetic moment $\mathbf{m}_i$
\begin{gather} \label{electron spin}
    \mathbf{m}_{i} = - g_{e} \mu_{B} \mathbf{s}_i = - \mathbf{s}_i
\end{gather}
interacts with the magnetic field of the cavity $\mathbf{B} \left( \mathbf{r}_i \right)$.
Here, $g_{e}$ is the electron \textit{g}-factor, $\mu_{B}$ is the Bohr magneton, and $\mathbf{s}_i$ is the electron spin operator associated with the electron i. In the presence of a homogeneous external magnetic field $\mathbf{B}_{ext}$ described by the external vector potential $\mathbf{A}_{ext}$, and nuclear magnetic moments $\mathbf{M}_{K}$ that give rise to the vector potential $\mathbf{A}_{n}$, Eq. \ref{H_minimal_coupling} may be written as
\begin{gather} \label{H_minimal_coupling_B}
    \begin{split}
        {H}_{mc}(\mathbf{B}_{ext}, \mathbf{M}) & = \frac{1}{2} \sum_{i} \bm{\pi}^2_{i}  - \sum_{iK} \frac{Z_{K}}{r_{iK}} + \frac{1}{2} \sum_{i \not = j} \frac{1}{r_{ij}} + \frac{1}{2} \sum_{K \not= L} \frac{Z_{K} Z_{L}}{R_{KL}} + \sum_{\mathbf{k}\lambda} \omega_{\mathbf{k}} b^{\dagger}_{\mathbf{k}\lambda} b_{\mathbf{k}\lambda} \\ & - \sum_{i} \mathbf{m}_{i} \cdot \mathbf{B}_{tot}(\mathbf{r}_{i}) - \sum_{K} \mathbf{M}_{K} \cdot \mathbf{B}_{tot}(\mathbf{R}_{K})
    \end{split}
\end{gather}
where $\mathbf{r}_i$ and $\mathbf{R}_K$ indicate the positions of the electron i and the nucleus K, respectively. Here, we refer collectively to the
magnetic moments by $\mathbf{M} = \{ \mathbf{M}_{K}\}$. In Eq. \ref{H_minimal_coupling_B}, the kinetic momentum operator now includes the following vector potential
\begin{gather} \label{vector_pot_B}
\begin{split}
    \mathbf{A}_{tot} \left( \mathbf{r}_{i} \right) & = \mathbf{A}\left( \mathbf{r}_{i} \right) + \mathbf{A}_{ext} \left( \mathbf{r}_{i} \right) + \mathbf{A}_{n} \left( \mathbf{r}_{i} \right) 
\end{split}.
\end{gather}
The vector potential associated with the static magnetic field is
\begin{gather}
\begin{split}
    \mathbf{A}_{ext} \left( \mathbf{r}_{i} \right) = \frac{\mathbf{B}_{ext} \times \mathbf{r}_{iO}}{2},
\end{split}
\end{gather}
where $r_{iO} = |\mathbf{r}_{i} - \mathbf{R}_{O}|$ is the distance between the electron i and the gauge origin $\mathbf{O}$. This term introduces a gauge origin dependence in the Hamiltonian that vanishes in the limit of a complete orbital basis \cite{helgaker1999}. With a truncated orbital basis, gauge origin independence is no longer guaranteed. To overcome this problem, we employed London Atomic Orbitals (LAOs) \cite{london1937}, as they have been extensively used in gauge origin-independent calculations of molecular magnetic properties \cite{helgaker1999,helgaker2012recent,aastrand1996magnetizabilities,ruud1993hartree,ruud1998hartree}. In Eq. \ref{vector_pot_B}, the vector potential from the nuclear magnetic moments is given by
\begin{gather} \label{A_nuc_pot}
    \mathbf{A}_{n} \left( \mathbf{r}_{i} \right) = \frac{1}{c^2} \sum_{K} \frac{\mathbf{M}_{K} \times \mathbf{r}_{iK}}{r_{iK}^3},
\end{gather}
where $r_{iK} = |\mathbf{r}_{i} - \mathbf{R}_{K}|$ is the distance between the electron i and the nucleus K. Note that Eq. \ref{A_nuc_pot} is invariant with respect to the choice of the origin. The curl of the vector potential in Eq. \ref{vector_pot_B} gives the total magnetic field
\begin{gather}
\begin{split}
    \mathbf{B}_{tot} \left( \mathbf{r}_i \right) & = \nabla_i \times \mathbf{A}_{tot} \left( \mathbf{r}_{i} \right)
\end{split},
\end{gather}
which is expressed as the sum of the different contributions
\begin{gather} \label{B_tot}
\begin{split}
    \mathbf{B}_{tot} \left( \mathbf{r}_i \right) & = \mathbf{B} \left( \mathbf{r}_i \right) + \mathbf{B}_{ext} \left( \mathbf{r}_i \right) + \mathbf{B}_{n} \left( \mathbf{r}_i \right) \\
    & = i \sum_{\mathbf{k} \lambda} \mathcal{A}_{\mathbf{k}} \left( \mathbf{k} \times \bm{\epsilon}_{\lambda} \right) ( b_{\mathbf{k} \lambda} e^{i \mathbf{k} \cdot \mathbf{r}_i} - b_{\mathbf{k} \lambda}^{\dagger} e^{-i \mathbf{k} \cdot \mathbf{r}_i} ) + \mathbf{B}_{ext} \\
    & + \frac{1}{c^2} \sum_{K} \left( \frac{3 \left( \mathbf{r}_{iK} \cdot \mathbf{M}_{K} \right) \mathbf{r}_{iK} - r_{iK}^2 \mathbf{M}_{K}}{r_{iK}^{5}}   + \frac{8 \pi}{3} \delta \left( \mathbf{r}_{iK} \right) \mathbf{M}_{K} \right).
\end{split}
\end{gather}
The interaction between the nuclear spin degrees of freedom with the total magnetic field is described through the Zeeman interactions in Eq. \ref{H_minimal_coupling_B}, where the nuclear magnetic moment $\mathbf{M}_{K}$ is
\begin{gather} \label{nuclear spin}
    \mathbf{M}_{K} = g_{K} \mu_{N} \mathbf{I}_{K} = \gamma_{K} \mathbf{I}_{K},
\end{gather}
where $g_{K}$ is the nuclear g-factor, $\mu_{N}$ is the nuclear magneton, $\gamma_{K}$ is the magnetogyric ratio, and $\mathbf{I}_{K}$ is the nuclear spin operator associated with the nucleus K.

\subsection{Cavity field approximation} \label{cav_app}
As a first approximation, the exact magnetic field $\mathbf{B}\left( \mathbf{r}_i \right)$ that enters the Hamiltonian in Eq. \ref{H_minimal_coupling_B} is written as
\begin{gather} \label{magnetic_field_app}
\begin{split}
    \mathbf{B} \left( \mathbf{r}_{i} \right) & = i \sum_{\mathbf{k} \lambda} \mathcal{A}_{\mathbf{k}} \left( \mathbf{k} \times \bm{\epsilon}_{\lambda} \right) \left( b_{\mathbf{k} \lambda} - b_{\mathbf{k} \lambda}^{\dagger} \right)
\end{split},
\end{gather}
where we have set $\text{exp} \left( \pm i \mathbf{k} \cdot \mathbf{r} \right) = 1$. Note that this approximation does not correspond to the commonly used dipole approximation where $|\mathbf{k}| = 0$, resulting in neglecting the magnetic contributions to the cavity field \cite{castagnola2023}. The associated vector potential is
\begin{gather}
\begin{split} \label{multipolar_A}
    \mathbf{A} \left( \mathbf{r}_{i} \right) &=  \sum_{\mathbf{k} \lambda} \mathcal{A}_{\mathbf{k}} \bm{\epsilon}_{\lambda} \left( b_{\mathbf{k} \lambda} + b_{\mathbf{k} \lambda}^{\dagger} \right) - \frac{i}{2} \sum_{\mathbf{k} \lambda} \mathcal{A}_{\mathbf{k}} \left( \mathbf{r}_{i} \times \left( \mathbf{k} \times \bm{\epsilon}_{\lambda} \right) \right) \left( b_{\mathbf{k} \lambda} - b_{\mathbf{k} \lambda}^{\dagger} \right)
\end{split},
\end{gather}
where the first term is the cavity vector potential in the dipole approximation and the second term gives rise to the cavity magnetic field in Eq. \ref{magnetic_field_app}. The corresponding electric field is given by
\begin{gather} \label{electric_field_app}
    \mathbf{E} \left( \mathbf{r}_{i} \right) = i \sum_{\mathbf{k} \lambda} \omega_{\mathbf{k}} \mathcal{A}_{\mathbf{k}} \bm{\epsilon}_{\lambda} \left( 
    b_{\mathbf{k} \lambda} - b^{\dagger}_{\mathbf{k} \lambda} \right) + \frac{1}{2} \sum_{\mathbf{k} \lambda} \omega_{\mathbf{k}} {\mathcal{A}_{\mathbf{k}}} \left( \mathbf{r}_i \times \left( \mathbf{k} \times \bm{\epsilon}_{\lambda} \right) \right) \left( b_{\mathbf{k} \lambda} + b^{\dagger}_{\mathbf{k} \lambda} \right).
\end{gather}
It is noteworthy that Eq. \ref{magnetic_field_app} and Eq. \ref{electric_field_app} fulfill the Maxwell's equations except for the Ampère-Maxwell's law \cite{jackson1977classical}
\begin{gather} \label{amp_law}
  \nabla \times \mathbf{B} = \frac{1}{c^2} \frac{\partial \mathbf{E}}{\partial t},
\end{gather}
as for the standard dipole approximation \cite{castagnola2023}. In fact, in Eq. \ref{amp_law}, the left-hand side is zero whereas the right-hand side does not vanish because of the time dependence of the photon operators in Eq. \ref{electric_field_app}. Further details related to this issue are reported in the Supplementary Information. Despite this limitation, using the cavity vector potential in Eq. \ref{multipolar_A} allows us to include the cavity magnetic dipole interaction terms. The total vector potential now takes the form
\begin{gather} \label{approx_A_tot}
\begin{split}
    \mathbf{A}_{tot} \left( \mathbf{r}_i \right) &= \sum_{\mathbf{k} \lambda} \mathcal{A}_{\mathbf{k}} \bm{\epsilon}_{\lambda} \left( b_{\mathbf{k} \lambda} + b_{\mathbf{k} \lambda}^{\dagger} \right) - \frac{i}{2} \sum_{\mathbf{k} \lambda} \mathcal{A}_{\mathbf{k}} \left( \mathbf{r}_{i} \times \left( \mathbf{k} \times \bm{\epsilon}_{\lambda} \right) \right) \left( b_{\mathbf{k} \lambda} - b_{\mathbf{k} \lambda}^{\dagger} \right) \\
    & + \frac{\mathbf{B}_{ext} \times \mathbf{r}_i}{2} + \frac{1}{c^2} \sum_{K} \left( \frac{\mathbf{M}_{K} \times \mathbf{r}_{i K}}{r_{i K}^3} \right).  
\end{split}
\end{gather}
The length gauge form of the Hamiltonian is obtained from Eq. \ref{approx_A_tot} by applying the following transformation
\begin{gather} \label{lengaugtransf}
\begin{split} 
    \mathbf{U} &= \text{exp} \left( i 
 \sum_{j} \mathbf{A}\left( 0 \right) \cdot \mathbf{r}_j \right) = \text{exp} \left( i  \sum_{j} \sum_{\mathbf{k} \lambda} \mathcal{A}_{\mathbf{k}} \left(\bm{\epsilon}_{\lambda} \cdot \mathbf{r}_{j} \right) \left(  b_{\mathbf{k} \lambda} + b_{\mathbf{k}  \lambda}^{\dagger} \right)  \right)
\end{split}.
\end{gather}
The transformed conjugate momentum $\bm{\pi}_i = \mathbf{p}_i + \mathbf{A}_{tot} \left( \mathbf{r}_i \right) $ then becomes
\begin{gather} \label{transformed_pi}
\begin{split}
    \mathbf{U} \bm{\pi}_i \mathbf{U}^{\dagger} &=  \mathbf{p}_i - \frac{1}{2} \sum_{ \mathbf{k} \lambda} \mathcal{A}_{\mathbf{k}} \left( \mathbf{r}_i \times \left( \mathbf{k} \times  \bm{\epsilon}_{\lambda} \right) \right) \left(  b_{\mathbf{k} \lambda} + b_{\mathbf{k} \lambda}^{\dagger} \right) \\
    & + 
    \frac{\mathbf{B}_{ext} \times \mathbf{r}_i}{2} + \frac{1}{c^2} \sum_{K} \left( \frac{\mathbf{M}_{K} \times \mathbf{r}_{i K}}{r_{i K}^3} \right).
\end{split}
\end{gather}
Here, we assumed that the cavity has at least two modes with wave vectors $\mathbf{k}$ and $-\mathbf{k}$, respectively. Therefore, the two-electron terms arising from the transformation of the total vector potential in Eq. \ref{approx_A_tot} vanish. Using the conjugate momentum in Eq. \ref{transformed_pi} and applying the unitary transformation
\begin{gather} \label{change_phase}
    \mathbf{V} = \text{exp} \left( i \frac{\pi}{2} \sum_{\mathbf{k} \lambda} b^{\dagger}_{\mathbf{k} \lambda} b_{\mathbf{k} \lambda} \right)
\end{gather}
we obtain the following Hamiltonian 
\begin{gather} \label{multipolar_H}
    \begin{split}
        H_{cav} \left(\mathbf{B}_{ext}, \mathbf{M}_{K}\right) &=  H (\mathbf{B}_{ext}, \mathbf{M}) \\
        & - \frac{1}{2}  \sum_{i} \sum_{\mathbf{k} \lambda} \mathcal{A}_{\mathbf{k}} \mathbf{l}_i \cdot \left( \mathbf{k} \times \bm{\epsilon}_{\lambda}  \right) \left( b_{\mathbf{k}\lambda} + b^{\dagger}_{\mathbf{k}\lambda} \right) \\
        & - \frac{1}{4} \sum_{i} \sum_{\mathbf{k}\lambda} \mathcal{A}_{\mathbf{k}} \left( 
        \mathbf{B}_{ext} \times \mathbf{r}_i \right) \cdot \left( \mathbf{r}_i \times \left( \mathbf{k} \times \bm{\epsilon}_{\lambda} \right) \right) \left( b_{\mathbf{k}\lambda} + b^{\dagger}_{\mathbf{k}\lambda} \right) \\
        & - \frac{1}{2 c^2} \sum_{i K} \sum_{\mathbf{k}\lambda} \mathcal{A}_{\mathbf{k}} \frac{\left( \mathbf{M}_{K} \times \mathbf{r}_{iK} \right) \cdot  \left( \mathbf{r}_i \times \left( \mathbf{k} \times \bm{\epsilon}_{\lambda} \right) \right)   }{r_{iK}^3}  \left( b_{\mathbf{k}\lambda} + b^{\dagger}_{\mathbf{k}\lambda} \right) \\
        & + \frac{1}{8} \sum_{i} \sum_{\mathbf{k}\lambda}\sum_{ \mathbf{k}'\lambda'} \mathcal{A}_{\mathbf{k}} \mathcal{A}_{\mathbf{k}'} \left( \mathbf{r}_i \times \left( \mathbf{k} \times \bm{\epsilon}_{\lambda} \right) \right) \cdot \left( \mathbf{r}_i \times \left( \mathbf{k}' \times \bm{\epsilon}_{\lambda'} \right) \right) \left( b_{\mathbf{k}\lambda} + b^{\dagger}_{\mathbf{k}\lambda} \right) \left( b_{\mathbf{k}'\lambda'} + b^{\dagger}_{\mathbf{k}'\lambda'} \right) \\
        &- \sum_{i} \sum_{\mathbf{k}\lambda} \mathcal{A}_{\mathbf{k}} \mathbf{m}_{i} \cdot \left( \mathbf{k} \times \bm{\epsilon}_{\lambda} \right) \left( b_{{\mathbf{k}\lambda}} + b^{\dagger}_{{\mathbf{k}\lambda}} \right) - \sum_{K} \sum_{\mathbf{k}\lambda} \mathcal{A}_{\mathbf{k}} \mathbf{M}_{K} \cdot \left( \mathbf{k} \times \bm{\epsilon}_{\lambda} \right)  \left( b_{{\mathbf{k}\lambda}} + b^{\dagger}_{{\mathbf{k}\lambda}} \right), 
    \end{split}
\end{gather}
where $\mathbf{l}_i = \mathbf{r}_{iO} \times \mathbf{p}_i$ is the angular momentum. Here, $H (\mathbf{B}_{ext}, \mathbf{M})$ represents the dipolar Hamiltonian that will be examined in the following Section. Notice that we have introduced new interaction terms that couple the external magnetic field, the electron spin, and the nuclear magnetic moments with the magnetic field of the cavity.

\subsection{Dipolar Hamiltonian} \label{dip_H}
To further simplify the Hamiltonian in Eq. \ref{multipolar_H}, we apply the dipole approximation by assuming that the relevant electromagnetic modes have a wavelength much larger than the characteristic lengths of the molecules. By setting $|\mathbf{k}| = 0$, we obtain the dipole Hamiltonian in the length gauge representation
\begin{gather} \label{qed_nmr_H_transf_imp}
    \begin{split}
        H (\mathbf{B}_{ext}, \mathbf{M}) &= H_{\text{PF}} + \mathbf{p} \cdot \mathbf{A}_{ext}\left( \mathbf{r} \right) + \mathbf{p} \cdot \mathbf{A}_{n}\left( \mathbf{r} \right) + \frac{\mathbf{A}_{ext} \left( \mathbf{r} \right)^2}{2} + \mathbf{A}_{ext} \left( \mathbf{r} \right) \mathbf{A}_n \left( \mathbf{r} \right) + \frac{\mathbf{A}_n \left( \mathbf{r} \right)^2}{2} \\
        & - \sum_{i} \mathbf{m}_{i} \cdot \left( \mathbf{B}_{ext} + \mathbf{B}_{n} \left( \mathbf{r}_{i} \right) \right) - \sum_{K} \mathbf{M}_{K} \cdot \left( \mathbf{B}_{ext} + \mathbf{B}_{n} \left( \mathbf{R}_{K} \right) \right).
    \end{split}
\end{gather}
Here, $H_{\text{PF}}$ is the standard Pauli-Fierz Hamiltonian \cite{castagnola2023}
\begin{gather} \label{H_PF}
    H_{\text{PF}} = H_{e} + \sum_{\alpha} \omega_{\alpha} b^{\dagger}_{\alpha} b_{\alpha} - \sum_{\alpha} \sqrt{\frac{\omega_{\alpha}}{2}} \left( \bm{\lambda}_{\alpha} \cdot \mathbf{d} \right) \left( b_{\alpha} + b_{\alpha}^{\dagger} \right) + \frac{1}{2} \sum_{\alpha}  \left( \bm{\lambda}_{\alpha} \cdot \mathbf{d} \right)^2
\end{gather}
where $H_{e}$ is the standard electronic Hamiltonian, $\mathbf{d}$ is the total dipole moment, and $\bm{\lambda}_{\alpha}$ the polarization vector, which are indicated as
\begin{gather}
    \mathbf{d} = - \sum_i \mathbf{r}_i + \sum_{K} Z_{K} \mathbf{R}_{K} \\
    \quad \bm{\lambda}_{\alpha} = \sqrt{\frac{2 \pi}{ \varepsilon_r V_{\alpha}}} \bm{\epsilon}_{\alpha},
\end{gather}
respectively. In Eq. \ref{H_PF}, we introduced $\alpha$ to denote the photonic mode defined by the wave vector $\mathbf{k}$ and polarization $\bm{\epsilon}_{\lambda}$. It is important to note that the Hamiltonian in Eq. \ref{qed_nmr_H_transf_imp} depends also on the choice of the origin of the multipole expansion. However, the origin invariance can be explicitly imposed by a suitable unitary transformation, as shown in Ref. \cite{Tor2020}. In the second quantization formalism the Hamiltonian in Eq. \ref{qed_nmr_H_transf_imp} may be written as
\begin{gather} \label{H_2q}
\begin{split}
    {H}(\mathbf{B}_{ext}, \mathbf{M}) &= \sum_{pq} \tilde{h}_{pq} E_{pq} + \frac{1}{2} \sum_{pqrs} \tilde{g}_{pqrs} \left( E_{pq} E_{rs} - \delta_{qr} E_{ps} \right) + \sum_{pq} V^{t}_{pq} \mathbf{T}_{pq} \\
    & + \sum_{\alpha}  \omega_{\alpha} b^{\dagger}_{\alpha} b_{\alpha} - \sum_{pq} \sum_{\alpha} \sqrt{\frac{\omega_{\alpha}}{2}} \left( \bm{\lambda}_{\alpha} \cdot \mathbf{d} \right)_{pq} \left( b_{\alpha} + b_{\alpha}^{\dagger} \right) E_{pq} \\
    & - \sum_{K} \mathbf{M}_{K} \cdot \left( \mathbf{B}_{ext} + \mathbf{B}_{n} \left( \mathbf{R}_{K} \right) \right),
\end{split}
\end{gather}
where p, q, r, and s denote the molecular orbitals. In Eq. \ref{H_2q}, we have introduced the singlet excitation operators
\begin{gather}
    E_{pq} = \sum_{\sigma} a^{\dagger}_{p\sigma} a_{q\sigma} 
\end{gather}
and the triplet excitation operators, which in the Cartesian representation read as
\begin{gather}
    \mathbf{T}_{pq} = \begin{bmatrix}
     T_{pq}^{x} \\ 
     T_{pq}^{y} \\ 
     T_{pq}^{z} \\ 
    \end{bmatrix} = \begin{bmatrix}
     \frac{1}{2} \left( a^{\dagger}_{p\sigma} a_{q\tau} + a^{\dagger}_{p\tau} a_{q\sigma} \right) \\ 
     \frac{1}{2i} \left( a^{\dagger}_{p\sigma} a_{q\tau} - a^{\dagger}_{p\tau} a_{q\sigma} \right) \\ 
     \frac{1}{2} \left( a^{\dagger}_{p\sigma} a_{q\sigma} - a^{\dagger}_{p\sigma} a_{q\sigma} \right)         
    \end{bmatrix}.
\end{gather}
Here, we have assumed that the creation and annihilation operators do not depend on the magnetic field $\mathbf{B}_{ext}$ since we are interested in the calculation of the energy derivatives. However, in the calculation of properties that involve the overlap of wave functions at different values of $\mathbf{B}_{ext}$, one must consider this dependence \cite{helgaker_electronic_1991}. The one-electron integrals in Eq. \ref{H_2q} include the one-electron dipole self-energy contribution and the first- and second-order singlet corrections due to the external fields
\begin{gather} \label{one-el}
    \begin{split}
        \tilde{h}_{pq} &= \sum_{iK} \bra{\varphi_{p} \left(\mathbf{B}_{ext}\right)}  \frac{\mathbf{p}_{i}^{2}}{2} -  \frac{Z_{K}}{r_{iK}} \ket{\varphi_{q}\left(\mathbf{B}_{ext}\right)} \\
        & + \frac{1}{2} \sum_{r} \sum_{\alpha} \bra{\varphi_{p} \left(\mathbf{B}_{ext}\right)} \left( \bm{\lambda}_{\alpha} \cdot \mathbf{d} \right) \ket{\varphi_{r} \left(\mathbf{B}_{ext}\right)} \bra{\varphi_{r} \left(\mathbf{B}_{ext}\right)} \left( \bm{\lambda}_{\alpha} \cdot \mathbf{d} \right) \ket{\varphi_{q} \left(\mathbf{B}_{ext}\right)} \\
        & + \sum_{i} \bra{\varphi_{p} \left(\mathbf{B}_{ext}\right)} \frac{\mathbf{l}_i \cdot \mathbf{B}_{ext}}{2} \ket{\varphi_{q} \left(\mathbf{B}_{ext}\right)} + \frac{1}{c^2} \sum_{iK} \bra{\varphi_{p} \left(\mathbf{B}_{ext}\right)} \frac{\mathbf{M}_{K} \cdot \mathbf{l}_{iK}}{r_{iK}^{3}} \ket{\varphi_{q} \left(\mathbf{B}_{ext}\right)} \\
        & + \frac{1}{8} \sum_{i} \bra{\varphi_{p} \left(\mathbf{B}_{ext}\right)} \left( B^2 r_{iO}^2 - \left( 
        \mathbf{B}_{ext} \cdot \mathbf{r}_{iO} \right)^2 \right) \ket{\varphi_{q} \left(\mathbf{B}_{ext}\right)} \\
        & + \frac{1}{2c^2} \sum_{iK} \bra{\varphi_{p} \left(\mathbf{B}_{ext}\right)} \frac{ \left( \mathbf{B}_{ext} \cdot \mathbf{M}_{K} \right) \left( \mathbf{r}_{iO} \cdot \mathbf{r}_{iK} \right) - \left( \mathbf{B}_{ext} \cdot \mathbf{r}_{iK} \right) \left( \mathbf{M}_{K} \cdot \mathbf{r}_{iO} \right)}{r_{iK}^{3}} \ket{\varphi_{q} \left(\mathbf{B}_{ext}\right)}\\
        & + \frac{1}{2c^4} \sum_{i} \sum_{K>L} \bra{\varphi_{p} \left(\mathbf{B}_{ext}\right)} \frac{ \left( \mathbf{M}_{K} \cdot \mathbf{M}_{L} \right) \left( \mathbf{r}_{iK} \cdot \mathbf{r}_{iL} \right) - \left( \mathbf{M}_{K} \cdot \mathbf{r}_{iL} \right) \left( \mathbf{r}_{iK} \cdot \mathbf{M}_{L} \right)}{r_{iK}^{3} r_{iL}^{3}} \ket{\varphi_{q} \left(\mathbf{B}_{ext}\right)},
    \end{split}
\end{gather}
where $\mathbf{l}_{iK}
 = \mathbf{r}_{iK} \times \mathbf{p}_i$ is the angular momentum around the nucleus K. The two-electron integrals now also include the two-electron dipole self-energy contribution
\begin{gather} \label{two-el}
    \begin{split}
        \tilde{g}_{pqrs} &= \sum_{i\not=j} \bigg( ( {\varphi_{p}\left( \mathbf{r}_i, \mathbf{B}_{ext} \right)\varphi_{q}\left( \mathbf{r}_i, \mathbf{B}_{ext} \right)}| \frac{1}{r_{ij}} |{\varphi_{r}\left(\mathbf{r}_j, \mathbf{B}_{ext} \right)\varphi_{s}\left( \mathbf{r}_j, \mathbf{B}_{ext} \right) ) } \\
        & +  \sum_{\alpha}  ( {\varphi_{p} \left( \mathbf{r}_i, \mathbf{B}_{ext} \right)} | \left( \bm{\lambda}_{\alpha} \cdot \mathbf{d} \right) |{\varphi_{q}\left(\mathbf{r}_i, \mathbf{B}_{ext} \right)}) ({\varphi_{r} \left( \mathbf{r}_j, \mathbf{B}_{ext} \right)} | \left( \bm{\lambda}_{\alpha} \cdot \mathbf{d} \right) | {\varphi_{s}\left( \mathbf{r}_j, \mathbf{B}_{ext} \right)} ) \bigg)
    \end{split}
\end{gather}
and the integrals coming from the electron spin operator are
\begin{gather} \label{triplet}
    \begin{split}
        V^{t}_{pq} &= \mathbf{B}_{ext}^{T} \delta_{pq} - \frac{1}{c^2} \sum_{iK} \mathbf{M}_{K}^{T} \bra{\varphi_{p} \left(\mathbf{B}_{ext}\right)}  \frac{r_{iK}^2\mathbf{1} - 3 \mathbf{r}_{iK} \mathbf{r}_{iK}^{T}}{r_{iK}^{5}} + \frac{8 \pi}{3} \delta \left( \mathbf{r}_{iK} \right) \ket{\varphi_{q} \left(\mathbf{B}_{ext}\right)}. 
    \end{split}
\end{gather}
A better description of the interactions represented by the integrals in Eq. \ref{one-el}, \ref{two-el}, and \ref{triplet} will be given in the next Section. Note that in Eq. \ref{one-el}, \ref{two-el}, and \ref{triplet} we introduced the LAOs, which are defined as
\begin{gather} \label{LAOs}
    \omega_{\mu} \left( \mathbf{B}_{ext} \right) = \text{exp} \left(- \frac{i}{2}{\mathbf{B}_{ext} \times \left( \mathbf{R}_{M} - \mathbf{R}_{O} \right) \cdot \mathbf{r} }\right) \chi_{\mu} \left( \mathbf{r} - \mathbf{R}_{M} \right)
\end{gather}
where $\chi_{\mu}$ is an atomic orbital centered on nucleus M at position $\mathbf{R}_{M}$, to ensure a gauge-invariant description of the atomic system in finite basis set calculations.

\subsection{Derivatives of the dipolar Hamiltonian}
The Hamiltonian in Eq. \ref{H_2q} is valid at all values of $\mathbf{B}_{ext}$ and $\mathbf{M}$. We may now expand this Hamiltonian in $\mathbf{B}_{ext}$ and $\mathbf{M}$ around $\mathbf{B}_{ext}=\mathbf{0}$ and $\mathbf{M}=\mathbf{0}$
\begin{gather}
    H(\mathbf{B}_{ext}, \mathbf{M}) = H^{(0)} + H^{(1)} \begin{bmatrix}
        \mathbf{B}_{ext} \\
        \mathbf{M}
    \end{bmatrix} + \frac{1}{2} \begin{bmatrix}
        \mathbf{B}_{ext}^T \mathbf{M}^T
    \end{bmatrix} H^{(2)} \begin{bmatrix}
        \mathbf{B}_{ext} \\
        \mathbf{M}
    \end{bmatrix} + \dots
\end{gather}
where the indices in the parenthesis $\left( n \right)$ denote the $n$-th order derivative. The zeroth-order Hamiltonian corresponds to the Pauli-Fierz Hamiltonian reported in Eq. \ref{H_PF}. The first-order Hamiltonian represents the paramagnetic interactions
\begin{gather} \label{magnetic}
    \frac{dH}{d\mathbf{B}_{ext}} = \frac{d H^{(0)}}{d \mathbf{B}_{ext}} + \sum_{i} \frac{\mathbf{l}_i}{2} - \sum_{i} \mathbf{m}_{i} \\
    \label{nuclear}
    \frac{dH}{d\mathbf{M}_K} = \frac{1}{c^2} \sum_{iK} \frac{\mathbf{l}_{iK} }{r_{iK}^3} + \frac{1}{c^2} \sum_{iK} \frac{r_{iK}^2 \mathbf{m}_i - 3 \left( \mathbf{m}_i \cdot \mathbf{r}_{iK} \right) \mathbf{r}_{iK}}{r_{iK}^5}  - \frac{8 \pi}{3 c^2} \sum_{i} \delta\left( \mathbf{r}_{iK} \right) \mathbf{m}_{i}.
\end{gather}
In Eq. \ref{magnetic}, the first term arises from the dependence of the atomic orbitals on the static magnetic field, the second term couples the orbital motion and the static magnetic field, and the last term arises from the electronic Zeeman interaction. In Eq. \ref{nuclear}, the first term represents the paramagnetic spin-orbit coupling. The last two terms correspond to the spin-dipole and the Fermi-contact interactions, which couple the nuclear magnetic moments to the electron spin.
The second-order interaction terms read as
\begin{align}
    \frac{d^2 H}{d \mathbf{B}_{ext}^2} &= \frac{d^2 H^{(0)}}{d \mathbf{B}_{ext}^2} + \frac{1}{8} \sum_{i} r_{iO}^2 \mathbf{1} -  \mathbf{r}_{iO} \mathbf{r}_{iO}^{T} \\
    \frac{d^2 H}{d \mathbf{B}_{ext} d \mathbf{M}_{K}} &= - \mathbf{1} + \frac{1}{2 c^2} \sum_{i} \frac{\left( \mathbf{r}_{iO} \cdot \mathbf{r}_{iK} \right) \bm{1} - \mathbf{r}_{iK} \mathbf{r}_{iO}^{T} }{r_{iK}^3} \label{shielding} \\
    \frac{d^2 H}{d \mathbf{M}_{K} d \mathbf{M}_{L}} &= \mathbf{D}_{KL} + \frac{1}{2 c^4} \sum_{i} \frac{\left( \mathbf{r}_{iK} \cdot \mathbf{r}_{iL} \right) \bm{1} - \mathbf{r}_{iK} \mathbf{r}_{iL}^{T} }{r_{iK}^3 r_{iL}^3} \label{spin-spin}
\end{align}
which correspond to the common diamagnetic interactions \cite{helgaker1999,helgaker_electronic_1991}. The purely nuclear contribution in Eq. \ref{shielding} arises from the nuclear Zeeman interaction, while in Eq. \ref{spin-spin}, it originates from the classical dipolar interaction, where $\mathbf{D}_{KL}$ is
\begin{gather}
     \mathbf{D}_{KL} = \frac{1}{c^{2}} \frac{{R}_{KL}^2 - 3 \mathbf{R}_{KL} \mathbf{R}_{KL}^{T} }{{R}_{KL}^5} .
\end{gather}
To construct the field-dependent molecular orbitals we employed the symmetric orbital connection proposed by Helgaker and J\o rgensen \cite{helgaker_electronic_1991}. In this formalism, we require the MOs to stay orthonormal for any value of the perturbing field. The Hamiltonian in Eq. \ref{H_2q} may be expressed employing a set of orthonormalized molecular orbitals (OMOs), which are written as
\begin{gather} \label{OMO}
    \varphi_{p}\left(\mathbf{B}_{ext}\right) = \sum_{m}  S^{-\frac{1}{2}}_{pm}\left(\mathbf{B}_{ext}\right) \phi_{m}\left(\mathbf{B}_{ext}\right),
\end{gather}
where
\begin{gather}
    \phi_m\left(\mathbf{B}_{ext}\right) = \sum_{\mu} \omega_{\mu}\left(\mathbf{B}_{ext}\right) C_{\mu m}\left(\mathbf{B}_{ext}=\mathbf{0}\right) 
\end{gather}
are the so-called unmodified molecular orbitals (UMOs), obtained by combining London atomic orbitals using the zero-field coefficients. The OMOs in Eq. \ref{OMO} are such that their derivative with respect to the magnetic field is
\begin{gather} \label{orb_connections}
    \frac{\partial \tilde{h}_{pq}^{OMO}}{\partial \mathbf{B}_{ext}} = \frac{\partial \tilde{h}_{pq}^{UMO}}{\partial \mathbf{B}_{ext}} - \frac{1}{2} \big\{ \frac{\partial {S}_{pq}^{UMO}}{\partial \mathbf{B}_{ext}}, \tilde{h}  \big\}_{pq},
\end{gather}
where the curly brackets represent the one-index transformed integrals
\begin{gather}
    \left\{ A, B \right\}_{pq} = \sum_{r} \left( A_{pr} B_{rq} + A_{qr}^{*} B_{pr} \right)
\end{gather}
and similarly in the case of the two-electron integrals \cite{helgaker_molecular_1986}. Therefore, when the n-th order derivative of the Hamiltonian in Eq. \ref{H_2q} is required, the contribution from the reorthonormalization of the molecular orbitals must be included, as shown in Eq. \ref{orb_connections}. For more details about orbital connections, we refer to these extended discussions in the literature \cite{helgaker_electronic_1991, helgaker_analytical_1988, helgaker_molecular_1986, olsen_orbital_1995}.

\subsection{QED-HF magnetic properties} \label{qed_hf_mag_prop}
\label{QED-HF_sub}
In the QED-HF model, the wave function ansatz is formulated as
\begin{gather}
    \ket{\text{R}} = \ket{\text{HF}} \otimes \ket{\text{P}}.
\end{gather}
Here, $\ket{\text{HF}}$ represents a single Slater determinant, and $\ket{\text{P}}$ is
\begin{gather}
    \ket{\text{P}} = \sum_{\mathbf{n}} \prod_{\alpha} \left( b_{\alpha}^{\dagger} \right)^{n_{\alpha}} \ket{0} c_{\mathbf{n}}, 
\end{gather}
where $\ket{0}$ denotes the photonic vacuum state, and $c_{\mathbf{n}}$ are the coefficients describing the expansion of photon number states. In the absence of external fields, the energy can be minimized with respect to the photon coefficients for a given HF state. This can be achieved by diagonalizing the photonic Hamiltonian,
\begin{gather}
\begin{split}
    \mel{\text{HF} }{ H_{\text{PF}}}{\text{HF}} &= E_{HF} + \sum_{\alpha} \bigg( \omega_{\alpha} b_{\alpha}^{\dagger} b_{\alpha} - \sqrt{\frac{\omega_{\alpha}}{2}} \left( \bm{\lambda}_{\alpha} \cdot \expval{\mathbf{d}} \right) \left( b_{\alpha} + b_{\alpha}^{\dagger} \right) + \frac{1}{2} \expval{\left( \bm{\lambda}_{\alpha} \cdot \mathbf{d} \right)^{2}} \bigg)
\end{split}
\end{gather}
through a unitary coherent-state transformation \cite{Tor2020}
\begin{gather} \label{coherent_state}
    \mathbf{W} = \prod_{\alpha} \text{exp} \left( z_{\alpha} \left( b_{\alpha} - b_{\alpha}^{\dagger} \right) \right)
\end{gather}
where $z_{\alpha}$ is chosen as
\begin{gather}
    z_{\alpha} = \frac{\bm{\lambda}_{\alpha} \cdot \expval{\mathbf{d}}}{\sqrt{2 \omega_{\alpha}}},
\end{gather}
and $\expval{\mathbf{d}}$ is written as
\begin{gather}
    \expval{\mathbf{d}} = \mel{\text{HF} }{\mathbf{d}}{\text{HF}}.
\end{gather}
The orbitals in the HF reference are optimized with an orthogonal transformation, defined as $\text{exp} \left( - \kappa \right)$, where $\kappa$ is an antisymmetric one-electron operator. In the coherent-state basis, the reference wave function is written as
\begin{gather}
     \ket{\Psi}= \prod_{\alpha} \text{exp} \left( - z_{\alpha} \left( b_{\alpha} - b_{\alpha}^{\dagger} \right) \right) \text{exp} \left( -\kappa \right) \ket{\text{R}}
\end{gather}
allowing the energy calculation to remain invariant with respect to the choice of the origin, even for charged molecules. Consequently, the polaritonic properties obtained through analytical energy derivatives are independent of the multipole expansion origin. In the presence of the external fields, the QED-HF energy may be written as
\begin{gather}
    E \left( \mathbf{B}_{ext}, \mathbf{M} \right) = \mel{\Psi \left( \zeta \right)}{{H}  \left( \mathbf{B}_{ext}, \mathbf{M} \right)}{\Psi \left( \zeta \right)},
\end{gather}
where $\zeta$ represents the optimized values of both electronic and photonic parameters, that satisfy the variational condition
\begin{gather} \label{var_cond}
    \frac{\partial E\left( \mathbf{B}_{ext}, \mathbf{M}\right)}{\partial \zeta} = 0
\end{gather}
for all values of $\mathbf{B}_{ext}$ and $\mathbf{M}$. Note that Eq. \ref{var_cond} determines the implicit dependence of the parameters $\zeta$ on the perturbations $\mathbf{B}_{ext}$ and $\mathbf{M}$. In addition, as the QED-HF method is variational, we can employ the standard procedure for variational wave functions to derive the expression of the polaritonic properties as analytical derivatives of the energy \cite{helgaker_1984_second}. The magnetic properties can be defined via the second-order derivatives as \cite{helgaker1999}
\begin{align}
    \bm{\chi} = -\mu_{0}   \frac{\partial^2 E \left( \mathbf{B}_{ext}, \mathbf{M}\right)}{\partial \mathbf{B}_{ext}^2} -\mu_{0} \frac{\partial^2 E \left( \mathbf{B}_{ext}, \mathbf{M}\right)}{\partial \mathbf{B}_{ext} \partial \zeta} \frac{\partial \zeta}{\partial \mathbf{B}_{ext}} \label{magsus}\\
    \bm{\sigma}_{K} - \mathbf{1} =  \frac{\partial^2 E \left( \mathbf{B}_{ext}, \mathbf{M}\right)}{\partial \mathbf{B}_{ext} \partial \mathbf{M}_{K}} + \frac{\partial^2 E \left( \mathbf{B}_{ext}, \mathbf{M}\right)}{\partial \mathbf{M}_{K} \partial \zeta} \frac{\partial \zeta}{\partial \mathbf{B}_{ext}} \label{shi}\\
    \mathbf{K}_{KL} + \mathbf{D}_{KL} =  \frac{\partial^2 E \left( \mathbf{B}_{ext}, \mathbf{M}\right)}{\partial \mathbf{M}_{K} \partial \mathbf{M}_{L}} + \frac{\partial^2 E \left( \mathbf{B}_{ext}, \mathbf{M}\right)}{\partial \mathbf{M}_{K} \partial \zeta} \frac{\partial \zeta}{\partial \mathbf{M}_{L}} \label{spn}
\end{align}
where $\bm{\chi}$ is the magnetizability tensor and $\mu_0$ is the magnetic permeability of free space, $\bm{\sigma}_{K}$ is the nuclear shielding tensor referred to the nucleus $K$, and $\mathbf{K}_{KL}$ is the indirect nuclear spin-spin coupling tensor between the nuclei $K$ and $L$. These second-order derivatives require only the first-order parameters with respect to the magnetic field $\mathbf{B}_{ext}$
or the nuclear dipole moment $\mathbf{M}_{L}$. The first-order parameters are obtained from the variational condition Eq. \ref{var_cond}, and they read
\begin{align} \label{magnetic_resp_imp}
    \frac{\partial^2 E \left( \mathbf{B}_{ext}, \mathbf{M}\right)}{\partial \zeta^2} \frac{\partial \zeta}{\partial \mathbf{B}_{ext}} &= - \frac{\partial^2 E \left( \mathbf{B}_{ext}, \mathbf{M}\right)}{\partial \mathbf{B}_{ext} \partial \zeta} \\
    \frac{\partial^2 E \left( \mathbf{B}_{ext}, \mathbf{M}\right)}{\partial \zeta^2} \frac{\partial \zeta}{\partial \mathbf{M}_{K}} &= -\frac{\partial^2 E \left( \mathbf{B}_{ext}, \mathbf{M}\right)}{\partial \mathbf{M}_{K} \partial \zeta}. 
\end{align}
These linear systems of equations can be solved iteratively and they require the first-order derivative with respect to the perturbation of the polaritonic energy gradient and the polaritonic energy Hessian. To derive explicit expressions for the QED-HF energy derivatives, we first express the Hamiltonian in Eq. \ref{H_2q} in the coherent-state basis
\begin{gather}
\begin{split}
    & \mathbf{W} {H} \left( \mathbf{B}_{ext}, \mathbf{M}\right) \mathbf{W}^{\dagger} = \sum_{pq} \tilde{h}_{pq} E_{pq} + \frac{1}{2} \sum_{pqrs} \tilde{g}_{pqrs} \left( E_{pq} E_{rs} - \delta_{qr} E_{ps} \right) \\
    & + \sum_{pq} V^t_{pq} \mathbf{T}_{pq} + \sum_{\alpha} \omega_{\alpha} b^{\dagger}_{\alpha} b_{\alpha} - \sum_{pq} \sum_{\alpha} \sqrt{\frac{\omega_{\alpha}}{2}} \left( \bm{\lambda}_{\alpha} \cdot \left( \mathbf{d} - \expval{\mathbf{d}} \right)\right)_{pq} \left( b_{\alpha} + b_{\alpha}^{\dagger} \right) E_{pq} \\
    & - \sum_{pq} \sum_{\alpha} \left( \bm{\lambda}_{\alpha} \cdot \mathbf{d} \right)_{pq} \left( \bm{\lambda}_{\alpha} \cdot \expval{\mathbf{d}} \right) E_{pq} + \frac{1}{2} \sum_{\alpha} \left( \bm{\lambda}_{\alpha} \cdot \expval{\mathbf{d}} \right)^2.
\end{split}
\end{gather}
For the magnetizability and the nuclear shieldings, the first-order response to the magnetic field is required, which is described by the imaginary part of the first-order parameters. Therefore, the QED-HF wave function can be parameterized as follows
\begin{gather}
    \ket{\Psi} = \text{exp} \left( - {\Lambda}  \right) \ket{R},
\end{gather}
where the operator ${\Lambda}$ may be chosen as
\begin{gather} \label{mag_parameter}
    {\Lambda} = \sum_{n} \upsilon_{n} {\Upsilon}_{n}  = i \sum_{\alpha} \prescript{I}{}{\gamma}_{\alpha} \left( b_{\alpha}^{\dagger} + b_{\alpha}  \right) + i \sum_{p>q} \prescript{I}{}{\kappa}_{pq} E_{pq}^{+},
\end{gather}
and the operator $E_{pq}^{+}$ is given by
\begin{gather}
    E_{pq}^{+} =  E_{pq} + E_{qp}.
\end{gather}
Here, $\gamma_{\alpha}$ describes the response of the coherent state to the perturbations, whereas $\kappa_{pq}$ represents the response of the orbitals including only non-redundant parameters. Following the general theory presented in Ref. \cite{castagnola2023}, we may write Eq. \ref{magsus}, and Eq. \ref{shi} as
\begin{gather}
    \bm{\chi} = -\mu_{0} \mel{\Psi} {\frac{\partial^2 {H}}{\partial \mathbf{B}_{ext}^2}}{\Psi} -\mu_{0} \mel{\Psi}{[ \frac{\partial {\Lambda}}{\partial \mathbf{B}_{ext}}, \frac{\partial {H}}{\partial \mathbf{B}_{ext}}]}{\Psi} \label{mag_exp} \\
    \bm{\sigma}_{K} = \mel{\Psi}{\frac{\partial^2 {H}}{\partial \mathbf{M} \partial \mathbf{B}_{ext}}}{\Psi} + \mel{\Psi}{[ \frac{\partial {\Lambda}}{\partial \mathbf{B}_{ext}}, \frac{\partial {H}}{\partial \mathbf{M}_K}]}{\Psi} \label{shi_exp}
\end{gather}
and the response equations in Eq. \ref{magnetic_resp_imp} as
\begin{gather}
    \sum_{n} \mel{\Psi}{[{\Upsilon}_m,[{\Upsilon}_n, {H}]]}{\Psi} \frac{\partial \upsilon_{n}}{\partial \mathbf{B}_{ext}} = - \mel{\Psi}{[{\Upsilon}_{m},\frac{\partial {H}}{\partial \mathbf{B}_{ext}}]}{\Psi}.
\end{gather}
Note that the elements of the Hessian on the left-hand side vanish for the coupling between electronic and photonic degrees of freedom. Similarly, the right-hand side is zero for the photonic operators as no terms couple the magnetic field $\mathbf{B}_{ext}$ with photonic degrees of freedom in the Hamiltonian. Therefore, the linear system of equations reduces to
\begin{gather}
    \sum_{r>s} \mel{\Psi}{[E_{pq}^{+},[{E}_{rs}^{+}, {H}]]}{\Psi} \frac{\partial \prescript{I}{}{\kappa}_{rs}}{\partial \mathbf{B}_{ext}} = i \mel{\Psi}{[E_{pq}^{+},\frac{\partial {H}}{\partial \mathbf{B}_{ext}}]}{\Psi}
\end{gather}
where the dipole self-energy contribution enters both the left- and right-hand sides. In the case of the second-order properties in Eq. \ref{mag_exp}-\ref{shi_exp}, the only new contribution arises using London atomic orbitals. The derivatives of the one-electron integrals are
\begin{gather}
    \begin{split}
    \frac{\partial\tilde{h}_{pq}}{\partial\mathbf{B}_{ext}} &= \frac{\partial{h}_{pq}}{\partial\mathbf{B}_{ext}}  + \frac{1}{2} \sum_{\alpha r} \frac{\partial \left( \bm{\lambda}_{\alpha} \cdot {\mathbf{d}} \right)_{pr} }{\partial \mathbf{B}_{ext}}  \left( \bm{\lambda}_{\alpha} \cdot {\mathbf{d}} \right)_{rq} + \frac{1}{2} \sum_{\alpha r} \left( \bm{\lambda}_{\alpha} \cdot {\mathbf{d}} \right)_{pr} \frac{\partial \left( \bm{\lambda}_{\alpha} \cdot {\mathbf{d}} \right)_{rq}}{\partial \mathbf{B}_{ext}} \\
    &+ \frac{1}{2} \sum_{\alpha} \sum_{\nu \rho} \left( \bm{\lambda}_{\alpha} \cdot {\mathbf{d}} \right)_{p\nu} \frac{\partial S_{\nu \rho}^{-1}}{\partial\mathbf{B}_{ext}} \left( \bm{\lambda}_{\alpha} \cdot {\mathbf{d}} \right)_{\rho q} - \sum_{\alpha} \frac{\partial \left( \bm{\lambda}_{\alpha} \cdot {\mathbf{d}} \right)_{pq}}{\partial \mathbf{B}_{ext}}  \left( \bm{\lambda}_{\alpha} \cdot \expval{\mathbf{d}} \right) \\
    & - \frac{1}{2} \big\{ \frac{\partial S}{\partial \mathbf{B}_{ext}}, \tilde{h} \big\}_{pq}
    \end{split},
\end{gather}
where the derivatives of the dipole operator and the inverse of the overlap matrix are also required. The derivatives of the two-electron integrals are given by
\begin{gather}
\begin{split}
    \frac{\partial \tilde{g}_{pqrs}}{\partial \mathbf{B}_{ext}} &= \frac{\partial {g}_{pqrs}}{\partial \mathbf{B}_{ext}} + \sum_{\alpha} \frac{\partial \left( \bm{\lambda}_{\alpha} \cdot \mathbf{d} \right)_{pq}} {\partial \mathbf{B}_{ext}} \left( \bm{\lambda}_{\alpha} \cdot \mathbf{d} \right)_{rs} + \left( \bm{\lambda}_{\alpha} \cdot \mathbf{d} \right)_{pq} \frac{\partial \left( \bm{\lambda}_{\alpha} \cdot \mathbf{d} \right)_{rs}}{\partial \mathbf{B}_{ext}} \\
    & - \frac{1}{2} \big\{ \frac{\partial S}{\partial \mathbf{B}_{ext}}, \tilde{g} \big\}_{pqrs}.
\end{split}
\end{gather}
Similarly, the second-order derivatives can be obtained using the technique described elsewhere \cite{helgaker_molecular_1986}.
To find explicit expressions for the indirect spin-spin coupling tensor $\mathbf{K}_{KL}$ in Eq. \ref{spn}, we need to change the wave function parameterization in Eq. \ref{mag_parameter}, as the nuclear perturbations involve triplet operators. The operator ${\Lambda}$ then may be written as
\begin{gather}
\begin{split}
    {\Lambda} &= i \sum_{\alpha} \prescript{I}{}{\gamma}_{\alpha} \left( b_{\alpha} + b_{\alpha}^{\dagger} \right) + \sum_{\alpha} \prescript{R}{}{\gamma}_{\alpha} \left( b_{\alpha} - b_{\alpha}^{\dagger} \right) \\
    & + i \sum_{p>q} \prescript{I}{}{\kappa}_{pq} E_{pq}^{+} + \sum_{p>q} \sum_{\xi} \prescript{R}{}{\kappa}_{pq}^{\xi} T_{pq}^{\xi -} + i \sum_{p>q} \sum_{\xi} \prescript{I}{}{\kappa}^{\xi}_{pq} T_{pq}^{\xi+} \\
\end{split}
\end{gather}
where the triplet operators are
\begin{gather}
    T_{pq}^{\xi -} = T_{pq}^{\xi} - T_{qp}^{\xi}, \quad
    T_{pq}^{\xi +} = T_{pq}^{\xi} + T_{qp}^{\xi}
\end{gather}
with $\xi$ labelling a Cartesian component between $x,y,z$. The expression for the indirect spin-spin coupling is obtained as
\begin{gather}
    \mathbf{K}_{KL} = \mel{\Psi} {\frac{\partial^2 {H}}{\partial \mathbf{M}_{K} \partial \mathbf{M}_{L}}}{\Psi} + \mel{\Psi}{[ \frac{\partial {\Lambda}}{\partial \mathbf{M}_K}, \frac{\partial {H}}{\partial \mathbf{M}_L}]}{\Psi}. \label{spn_exp}
\end{gather}
The evaluation of this property requires solving the nuclear response equations for the orbital response
\begin{align} \label{spn_resp}
\begin{split}
    \sum_{r>s} \mel{\Psi}{[E_{pq}^{+},[E_{rs}^{+}, {H}]]}{\Psi} \frac{\partial \prescript{I}{}{\kappa}_{rs}}{\partial \mathbf{M}_K} &= i\mel{\Psi}{[E_{pq}^{+},  \frac{\partial {H}}{\partial \mathbf{M}_K} ]}{\Psi} \\
    \sum_{r>s} \sum_{\xi} \mel{\Psi}{[T_{pq}^{\xi -},[T_{rs}^{\xi -}, {H}]]}{\Psi} \frac{\partial \prescript{R}{}{\kappa}^{\xi}_{rs}}{\partial \mathbf{M}_K} &= - \mel{\Psi}{[T_{pq}^{\xi -}, \frac{\partial {H}}{\partial \mathbf{M}_K} ]}{\Psi}
\end{split}
\end{align}
which now involve the dipole self-energy contributions to the Hessian on the left-hand side. The explicit expression of the indirect spin-spin coupling remains unchanged. However, the effects of the dipole self-energy are now included in the wave function response.

\section{Results and discussions}
The calculation of the HF and QED-HF magnetic properties have been implemented in a development version of the eT program \cite{folkestad2020t}. All molecular geometries used in this paper have been optimized using the ORCA software package \cite{neese2012orca} using a DFT-B3LYP level of theory and a def2-SVP basis set \cite{weigend_def2_svp_2005}. These geometries are available in the Supplementary Information. All calculations of the magnetic properties reported in this paper have been performed using an aug-cc-pVDZ basis set \cite{kendall_aug_basis_1992,woon_aug_basis_1993}. In the following sections, we present the QED-HF magnetizabilities for a range of hydrocarbons comparing them with the no-cavity HF values. Additionally, we report the effects of strong light-matter coupling on the aromaticity descriptors as the NICS and magnetizability exaltation used to study the reaction pathway of the acetylene trimerization to benzene in optical cavity.

\subsection{Modulation of magnetizabilities} \label{mag_hydro}
In this section, we explore the strong light-matter coupling effects on the magnetizabilities of the methane, ethylene, acetylene, and benzene molecules. We examined the effect of different polarization orientations and coupling strength on the isotropic magnetizabilities
\begin{gather}
    \chi_{iso} = \frac{\chi_{xx} + \chi_{yy} + \chi_{zz}}{3},
\end{gather}
where $\chi_{xx}$, $\chi_{yy}$, and $\chi_{zz}$ are the diagonal elements of the magnetizability tensor. The QED-HF calculations were conducted within an optical cavity with a frequency of 2.7 eV and a coupling strength of 0.1 a.u. The methane molecule was positioned with the carbon atom in the origin and the two couples of protons aligned along the x- and y-axis, respectively. The ethylene and acetylene were positioned within the cavity with the C-C bonds aligned along the x-axis, whereas the benzene molecule was oriented to lie in the xy-plane. The three different polarization orientations were chosen along the x-, y- and z-axis, as shown in Fig. \ref{fig:opt_cav}.
\begin{table}[H]
  \caption{Isotropic magnetizabilities for different polarization directions of the cavity field ($10^{-30}~\text{J}~\text{T}^{-2}$).}
  \label{tbl:qed_iso_mag}
  \begin{tabular}{ccccc}
    \hline
   molecule & HF  & QED-HF, x &  QED-HF, y  & QED-HF, z  \\
    \hline
    methane & -317.23 & -312.68 & -312.68 & -312.60 \\
    ethylene & -360.39 & -355.84 & -359.22 & -352.94 \\
    acetylene & -388.37 & -382.89 & -381.62 & -381.62 \\
    benzene & -991.77 & -990.50 & -989.90 & -980.39 \\
    \hline
  \end{tabular}
\end{table}
\begin{table}[H]
  \caption{Out-of-plane component of the magnetizabilities tensors for the benzene molecule ($10^{-30}~\text{J}~\text{T}^{-2}$).}
  \label{tbl:qed_ooc_ben_mag}
  \begin{tabular}{cc}
    \hline
     Method & $\chi_{zz}$ \\
    \hline
     HF & -1705.46 \\
     QED-HF, x & -1701.75 \\
     QED-HF, y & -1700.82 \\
     QED-HF, z & -1687.65 \\
    \hline
  \end{tabular}
\end{table}
A comparison between the HF and QED-HF isotropic magnetizabilities hydrocarbons is presented in Tab. \ref{tbl:qed_iso_mag}. For methane, the high degree of symmetry results in a change that is almost the same for all polarization orientations. The cavity induces different changes for acetylene when the polarization is oriented along the x-axis, as the $\sigma$ bonds are involved. For the other two polarization orientations, the cavity effects are comparable since the $\pi$ bonds are equally affected. In the case of benzene, shifts in the isotropic magnetizability can be explained by considering its aromatic character. To describe this feature, the out-of-plane component of the magnetizability tensor can be employed to examine the delocalization of the $\pi$ electrons \cite{lazzeretti2004}. The more delocalized the $\pi$ electrons, the higher the absolute value of the out-of-plane magnetizability. 
As shown in Tab. \ref{tbl:qed_ooc_ben_mag}, when the polarization lies along the plane of the molecule, it induces minor changes in the out-of-plane components, indicating that the $\pi$ electrons exhibit a relatively small response to the quantum electromagnetic field. However, when the polarization is orthogonal to the molecular plane, decreased delocalization occurs, resulting in a decrease (in absolute value) of the isotropic magnetizability. The cavity alters the distribution of the electron density over the aromatic ring leading to a decrease in the aromatic character of the molecule. This behavior has also been observed in the ethylene molecule, where the largest cavity effect emerges when the polarization is aligned with the $\pi$-bond orbitals.
\begin{figure}[ht]
    \begin{subfigure}{0.5\textwidth}
        \centering
        \caption{}
        \includegraphics[scale=0.5]{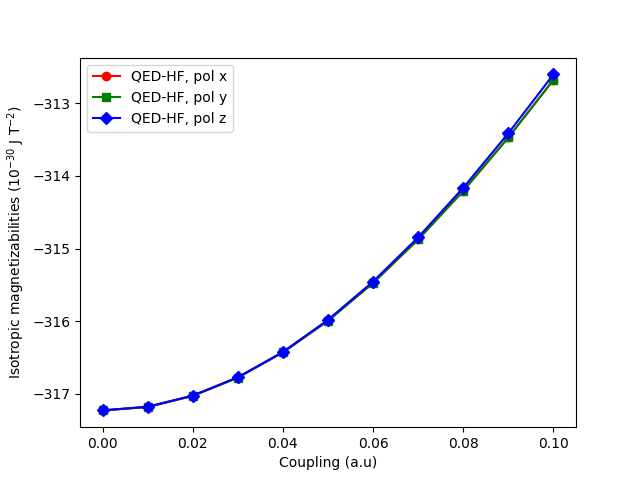}
        \label{fig:methane_mag}
    \end{subfigure}%
    \begin{subfigure}{0.5\textwidth}
        \centering
        \caption{}
        \includegraphics[scale=0.5]{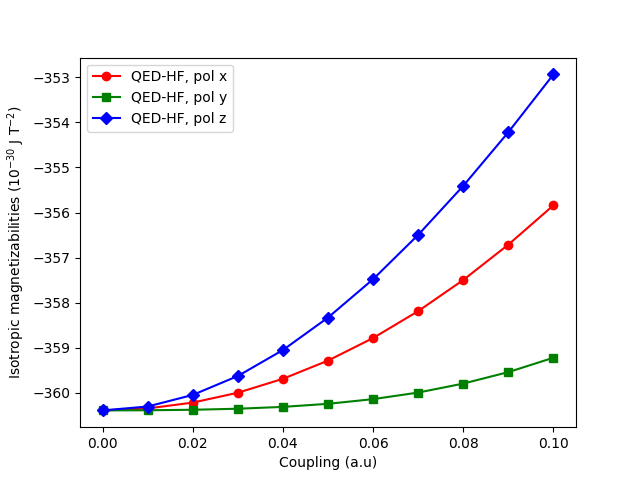}
        \label{fig:ethylene_mag}
    \end{subfigure}
    \begin{subfigure}{0.5\textwidth}
        \centering
        \caption{}
        \includegraphics[scale=0.5]{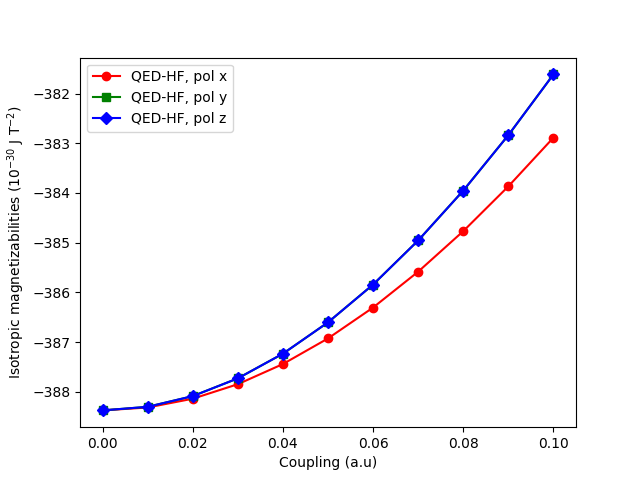}
        \label{fig:acetylene_mag}
    \end{subfigure}%
    \begin{subfigure}{0.5\textwidth}
        \centering
        \caption{}
        \includegraphics[scale=0.5]{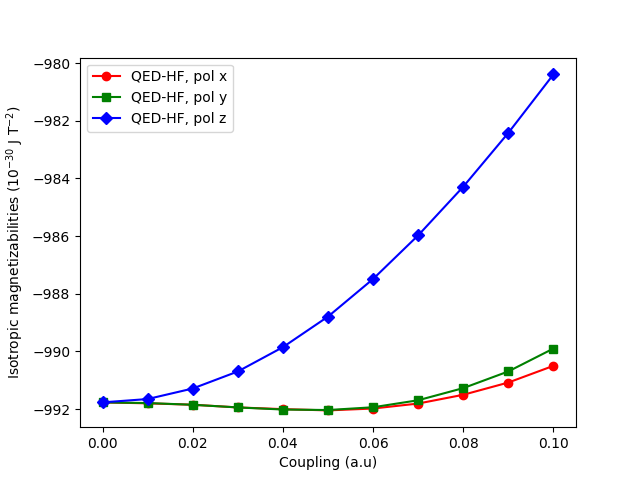}
        \label{fig:benzene_mag}
    \end{subfigure}
    \caption{Variation in the total isotropic magnetizability at different coupling values and polarization orientations for the methane (a), ethylene (b), acetylene (c), and benzene molecules (d).}
    \label{fig:mag_coupling_tot}
\end{figure}
In Fig. \ref{fig:mag_coupling_tot} we illustrate the isotropic magnetizability as a function of coupling strength showing the different polarization orientation effects for each hydrocarbon. Methane shows a consistent curve shape independently of the polarization orientation (Fig. \ref{fig:mag_coupling_tot}a). Acetylene shows a similar behavior mirroring the trend observed for methane except for the polarization oriented along the bond axis (Fig. \ref{fig:mag_coupling_tot}c). However, in the case of ethylene and benzene (Fig. \ref{fig:mag_coupling_tot}b, d), the changes with the coupling confirm that the polarization orthogonal to the molecular plane produces a larger effect in the isotropic magnetizability while the coupling is increasing, affecting largely the out-of-plane component of the magnetizability. Moreover, the benzene molecule shows quite peculiar behavior when polarization is oriented along the molecular plane. Indeed, for small values of coupling strength, the magnetizability slightly increases while for larger couplings starts to decrease (in absolute value). Additionally, the in-plane polarizations affect differently the molecular orbitals, leading to different values of magnetizability when the coupling is increased. It is worth mentioning that the HF model effectively reproduces experimental magnetizabilities values, as the contribution of the electron correlation to this property is usually small \cite{ruud1994hydrocarbons}. However, in the case of polaritons, the effects of electron-photon correlation could play a more important role on this property. Consequently, further investigations are necessary to elucidate these effects by using electron-photon correlated models, for instance, QED-CC \cite{Tor2020}.

\subsection{Modulation of aromaticity} \label{trim_cav}
\begin{figure}[H]
    \centering
    \includegraphics[scale = 0.4]{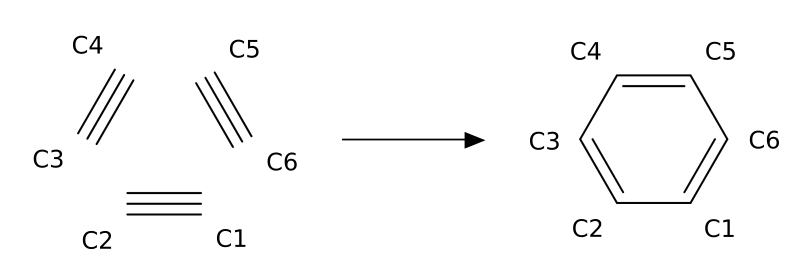}
    \caption{Schematic representation of the reaction mechanism for the trimerization of acetylene to benzene.}
    \label{fig:reaction}
\end{figure}

\begin{figure}[H]
    \begin{subfigure}{0.5\textwidth}
       \centering
        \caption{}
        \includegraphics[scale = 0.5]{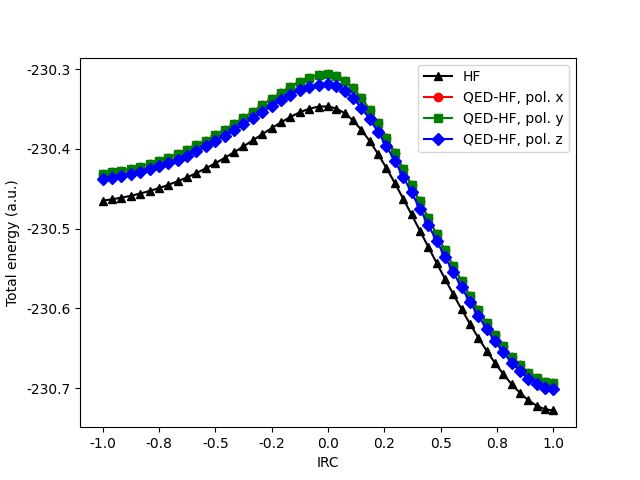}
        \label{fig:energies_irc}
    \end{subfigure}%
    \begin{subfigure}{0.5\textwidth}
        \centering
        \caption{}
        \includegraphics[scale = 0.5]{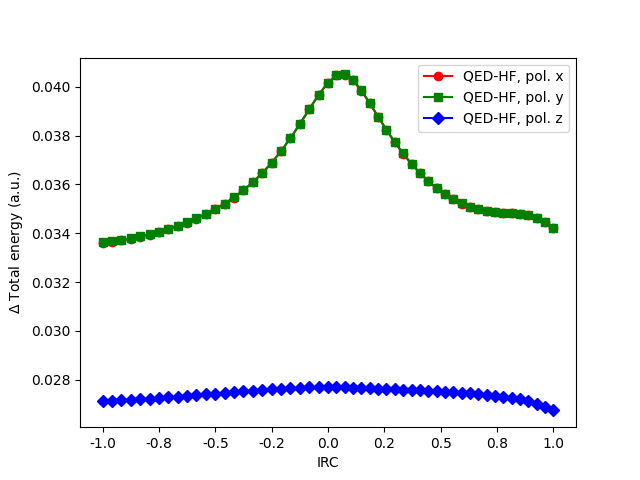}
        \label{fig:delta_energies_irc}
    \end{subfigure}
    \caption{Comparison of the total energy (a) and its differences (b) for the HF and QED-HF calculations with different polarization orientations along the IRC.}
    \label{fig:energy_trim_tot}
\end{figure}

\begin{figure}[H]
    \begin{subfigure}{0.5\textwidth}
        \centering
        \caption{}
        \includegraphics[scale=0.5]{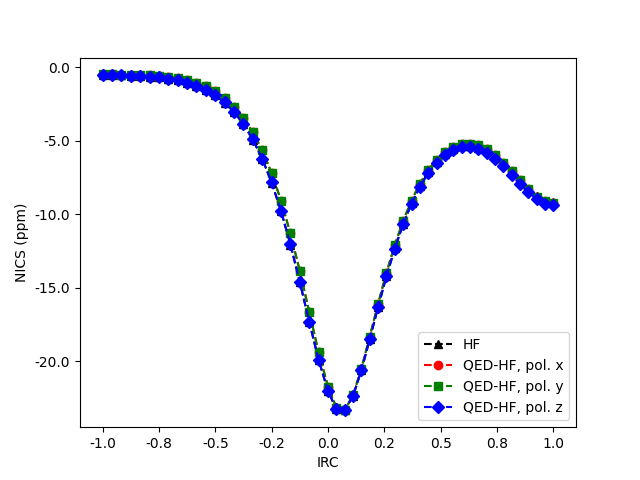}
        \label{fig:nics_irc}
    \end{subfigure}%
    \begin{subfigure}{0.5\textwidth}
        \centering
        \caption{}
        \includegraphics[scale=0.5]{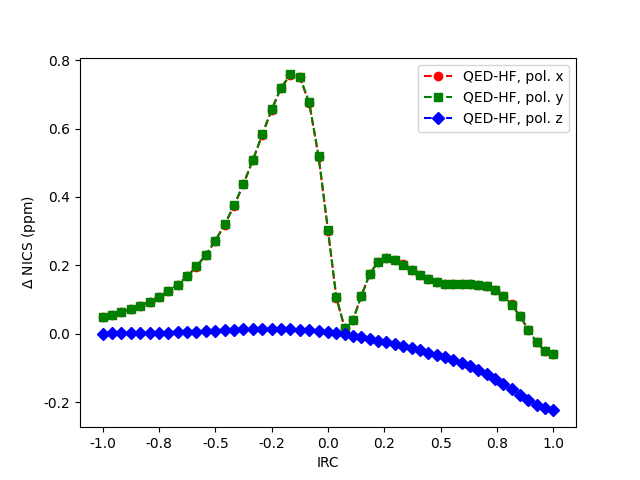}
        \label{fig:delta_nics_irc}
    \end{subfigure}
    \caption{Comparison of NICS$_{tot}$(0) (a) and its differences (b) for the HF and QED-HF calculations with different polarization orientations along the IRC.}
    \label{fig:nics_trim_tot}
\end{figure}

We conclude by investigating the quantum field effects on the trimerization of acetylene to benzene, represented in Fig. \ref{fig:reaction}. This reaction is an example of thermally allowed pericyclic reactions intensively studied in the past \cite{houk_trim_1979,ioffe_intra_1992,jiao_aroma_1998,cioslowski_trim_2000,remco_trime_2003} and takes place via a concerted pathway that passes through an aromatic transition state (TS). This transition state has been theoretically investigated analyzing various magnetic properties as $\prescript{1}{}{\text{H}}$ NMR chemical shifts \cite{jiao_nmr_trime_1993,jiao_magsus_trime_1994,jiao_mag_trim_1995}, magnetizability exaltation \cite{herges_magsus_trim_1994,jiao_electro_1994,jiao_electrostatic_au_1995,jiao_aroma_1998}, and nucleus independent chemical shift (NICS) \cite{remco_trime_2003}. 
In this work, we employed the nucleus-independent chemical shift at the ring center, known as NICS(0), and the magnetizability exaltation, as aromaticity descriptors \cite{poranne2015}. They are defined as
\begin{gather}
    \text{NICS(0)} = \frac{\sigma_{xx} + \sigma_{yy} + \sigma_{zz}}{3} \label{nics_0} \\
    \chi_{ex} = \chi_{iso}^{TS} - \chi_{iso}^{R}. \label{mag_ex}
\end{gather}
In Eq. \ref{nics_0}, $\sigma_{xx}$, $\sigma_{yy}$, and $\sigma_{zz}$ represent the diagonal components of the nuclear shielding tensor for a ghost atom placed in the center of mass of the molecule. In Eq. \ref{mag_ex}, $\chi_{iso}^{TS}$ is the isotropic magnetizability of the transition state and $\chi_{iso}^{R}$ is the magnetizability of the reactants. As the trimerization reaction is symmetry-allowed according to the Woodward and Hoffmann rules \cite{woodward_cons_1971}, employing a single determinant as an electronic wave function is sufficient for qualitatively describing the essential features of the reaction. To generate the reaction pathway we employed the intrinsic reaction coordinate (IRC) calculations using the ORCA software package \cite{neese2012orca} with the nudged elastic band and transition state optimization (NEB-TS) method \cite{asgeirsson_neb_ts_2021}. An atom-pairwise dispersion correction based on tight binding partial charges \cite{caldeweyher_d4_2017} has been also applied. These calculations were performed at the DFT-B3LYP/def2-SVP level of theory. The starting geometries of the reactants and products were taken from Ref. \cite{remco_trime_2003}. The reactant geometry is considered to have IRC = -1, the transition state has IRC = 0 by definition, and the equilibrium geometry of the product has IRC = 1. A set of 22 geometries has been computed from IRC = -1 to IRC = 0, and an additional 27 geometries from IRC = 0 to IRC = 1. The transition state has D$_{3h}$ symmetry with a single imaginary vibrational frequency at -616.7 \text{cm}$^{-1}$ and carbon-carbon separations of 1.23 \AA~ and 2.33 \AA. These findings are in line with a previous study by Jiao et al. \cite{jiao_aroma_1998}. Figure \ref{fig:energies_irc} shows the total energies obtained for HF and QED-HF. The QED-HF calculations were carried out with $\omega$ = 1.90 eV and $\lambda$ = 0.05 for different polarization directions along the x-, y-, and z-axis. The reactants and products were positioned in the xy plane. The HF reproduces well previous results \cite{jiao_aroma_1998}. Examining the potential energy surface along the IRC, a relatively flat region is observed from acetylene reactants to the transition state, followed by a steep descent to benzene. The QED-HF curves confirm the concerted and synchronous nature of the transition from reactants to products, even under the influence of a quantum electromagnetic field. However, Fig. \ref{fig:delta_energies_irc} reveals that the orientation of polarization influences differently the total energy along the reaction pathway. When polarization is orthogonal to the plane containing the reactants, a modest shift in total energy occurs. In contrast, for the in-plane polarizations, the effect of the quantum field intensifies as the transition state is approached. This effect could be attributed to the larger polarization of the orbitals, manifested through increased oscillations of the total electronic dipole around its mean value. The activation energies in Table \ref{tbl:energies} support and confirm the observed behavior.
\begin{table}[H]
\caption{Activation energies (kcal mol$^{-1}$) for the HF and QED-HF calculations with different polarization orientations.}
\label{tbl:energies}
\centering
\begin{tabular}{cc}
\hline
Method & Act. Energies \\
\hline
HF & 74.04 \\
QED-HF, x & 78.44 \\
QED-HF, y & 78.44 \\
QED-HF, z & 74.67 \\
\hline
\end{tabular}
\end{table}
The NICS(0) results are reported in Fig. \ref{fig:nics_irc}. The HF curve is in agreement with the findings of Remco et al. \cite{remco_trime_2003}. The negative values indicate the aromatic character of the transition state and the products. As suggested by the authors, the NICS(0) remains close to zero in the early stages of the reaction, decreases to a minimum immediately after reaching the transition state, and then raises again as the paratropic character increases. Finally, it decreases again due to the formation of the $\pi$ bonds. As the NICS(0) is calculated in the molecular plane, the TS NICS(0) is higher (in absolute value) due to the $\sigma$ electrons ring current, which is less intense in the case of benzene. In Fig. \ref{fig:delta_nics_irc}, we reported the differences in the NICS(0) between the QED-HF and HF results along the IRC. As observed, when the polarization is aligned with the z-axis the QED-HF remains equal to the HF. However, at the end of the reaction path, it increases (in absolute value) meaning that the polarization of the $\pi$ electrons due to the cavity increases the diatropic character. In the case of x- and y- polarization the QED-HF is lower than the HF (in absolute value) until the transition state is approached meaning that the diatropic character is decreased by the cavity. This behavior confirms that the polarizations within the molecular plane lead to a decrease in NICS(0) at the transition state. Consequently, the aromatic character is reduced by the cavity, resulting in a less stable transition state, as confirmed by the activation energy analysis. Subsequently, after reaching the transition state, the value converges towards the HF value to increase again later, where the paratropic character decreases (in absolute value). Finally, the QED-HF approaches the HF values at the end of the reaction pathway.
In Table \ref{tbl:magnet_ts} we reported the diagonal elements of the magnetizability tensors for the transition state obtained with HF and QED-HF methods. The HF results are in line with the aromaticity evaluation of the transition state reported by Jiao et al. \cite{jiao_aroma_1998}. This agreement persists in the QED-HF results. However, for the x- and y- polarization the out-of-plane component of the tensors shows a decrease compared to HF values. This behavior is in line with the NICS(0) results, suggesting a slightly decreased aromatic character of the transition state within the cavity. Moreover, the in-plane components are almost identical due to the high degree of symmetry of the transition state. On the contrary, the polarization along the z-axis produces a shift in all the diagonal components of the tensor, similar to what is observed for the total energies.
\begin{table}[H] 
  \centering
  \begin{subtable}{0.45\textwidth}
    \centering
    \caption{}
    \label{tbl:magnet_ts}
    \begin{tabular}{cccc}
      \hline
      Method & $\chi_{xx}$ & $\chi_{yy}$ & $\chi_{zz}$ \\
      \hline 
      \text{HF} & -993.11 & -993.03 & -1978.91 \\
      \text{QED-HF, x} & -994.52 & -988.02 & -1830.38 \\ 
      \text{QED-HF, y} & -988.10 & -994.93 & -1827.79 \\
      \text{QED-HF, z} & -974.99 & -974.95 & -1961.37 \\
      \hline
    \end{tabular}
  \end{subtable}
  \hspace{1cm}
  \begin{subtable}{0.45\textwidth}
    \centering
    \caption{}
    \begin{tabular}{cc}
      \hline
      Method & $\chi_{ex}$ \\
      \hline
      HF & -12.80 \\
      QED-HF, x & -10.65 \\
      QED-HF, y & -10.64 \\
      QED-HF, z & -12.78 \\
      \hline
    \end{tabular}
    \label{tab:lambda-values}
  \end{subtable}
  \caption{Diagonal elements of the transition state magnetizability tensors (a) and magnetizability exaltation values (b) for the HF and QED-HF calculations ($10^{-30}~\text{J}~\text{T}^{-2}$).}
\end{table}
The magnetizability exaltations are shown in Tab \ref{tab:lambda-values}. The negative values are in line with the aromatic character of the transition state. Moreover, the QED-HF results confirm that z-polarization merely causes a shift in values, as the obtained value aligns with the HF values. On the contrary, the x- and y-directions modify the features of the transition state, decreasing its aromatic character, as demonstrated also by the analysis of NICS(0).

\section{Conclusions} \label{conclusions}
In this work, we have developed \textit{ab initio} methods that explicitly include interactions with a static magnetic field and the nuclear spin degrees of freedom for molecular systems within an optical cavity. Firstly, we introduced a minimal coupling approach that completely describes these interactions. Subsequently, we presented a model that includes the cavity magnetic dipole interactions with an approximate description of the quantum electromagnetic field. Finally, we further simplified this Hamiltonian by applying the dipole approximation. We developed the first implementation at the QED-HF level for calculating magnetizability and nuclear shielding tensors. The obtained results for the magnetizability of hydrocarbons indicate significant effects induced by the cavity. Indeed, the isotropic magnetizability varies depending on the polarization orientation and the coupling strength. In aromatic compounds such as benzene, we observed that the predominant effect of the cavity occurs when the polarization is orthogonal to the molecular plane. This is confirmed by changes in the out-of-plane component of the magnetizability, which indicate a decreased delocalization of the $\pi$ electrons with a consequent alteration in the electron density distribution over the aromatic ring. Furthermore, we explored the effects of the optical cavity on aromaticity descriptors. The results obtained from the acetylene trimerization to benzene indicate that the cavity can modify the aromatic character of the transition state, as highlighted by NICS values and magnetizability exaltation. We demonstrated that when the polarization is oriented in the plane of molecules, there is an increase in the activation energy. This modification could lead to a shift in the equilibrium of the reaction, offering a way to govern the reaction pathway that involves aromatic transition states or intermediates, even if the effects induced by the cavity are small compared to the electron stabilization in aromatic systems. This study opens the possibility to further investigations on how molecular magnetic properties are influenced by the presence of a quantum electromagnetic field. Future analysis may include the electron-photon correlation in the calculation of such properties.

\section{Data and code availability}
The data and the code that support the ﬁndings of this study are available from the corresponding author upon reasonable request. Examples of the input files used to run the calculations are available in Ref. \cite{barlini_mag_prop_2024}.

\begin{acknowledgement}
The authors thank Matteo Rinaldi and Rosario Roberto Riso for their insightful advice and discussions. Alberto Barlini, Andrea Bianchi, and Henrik Koch acknowledge funding from the European Research Council (ERC) under the European Union's Horizon 2020 Research and Innovation Programme (grant agreement No. 101020016). Enrico Ronca acknowledges funding from the European Research Council (ERC) under the European Union's Horizon Europe Research and Innovation Programme (Grant n. ERC-StG-2021-101040197 – QED-SPIN). 
\end{acknowledgement}


\providecommand{\latin}[1]{#1}
\makeatletter
\providecommand{\doi}
  {\begingroup\let\do\@makeother\dospecials
  \catcode`\{=1 \catcode`\}=2 \doi@aux}
\providecommand{\doi@aux}[1]{\endgroup\texttt{#1}}
\makeatother
\providecommand*\mcitethebibliography{\thebibliography}
\csname @ifundefined\endcsname{endmcitethebibliography}  {\let\endmcitethebibliography\endthebibliography}{}

\clearpage
\includepdf[pages=-]{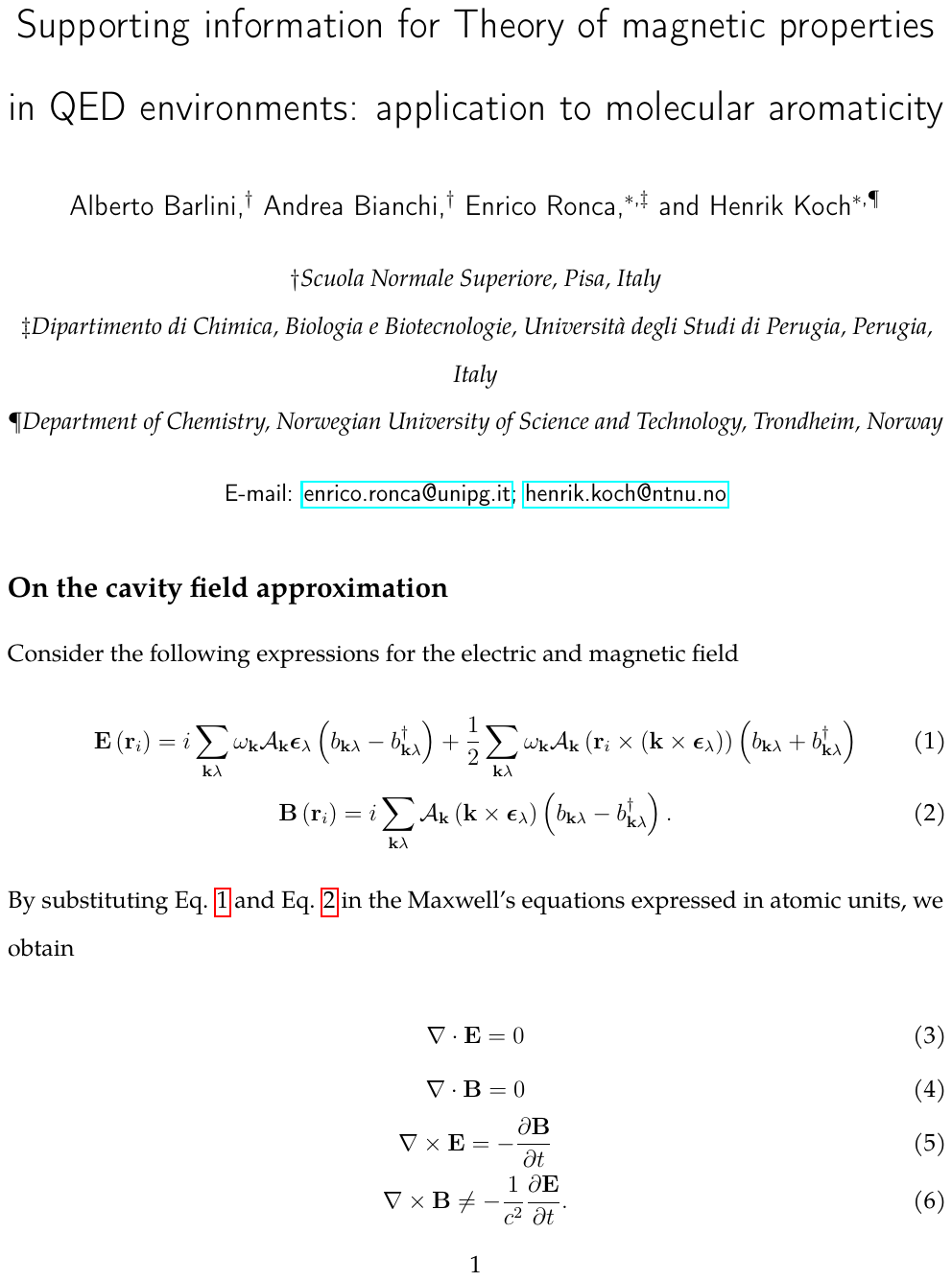}

\end{document}